\documentclass[]{aa}
\usepackage{graphicx}
\usepackage{amsmath,amssymb,amsbsy}
\usepackage{natbib}
\usepackage{txfonts}
\usepackage{latexsym}
\usepackage{rotating}
\usepackage{float}

\newcommand{\Halpha}{\ensuremath{\mathrm{H}\alpha}}
\newcommand{\zav}[1]{\left(#1\right)}
\newcommand{\kms}{\ensuremath{\mathrm{km}\,\mathrm{s}^{-1}}}

\newcommand\de{\text{d}}
\newcommand{\nahoda}{\ensuremath{\xi}}
\newcommand{\tnahoda}{\ensuremath{\tau_\nahoda}}
\newcommand{\deriv}[2]{\frac{\mathrm{d} #1}{\mathrm{d} #2}}
\newcommand{\rcl}{\ensuremath{r_\text{cl}}}
\newcommand{\rmin}{\ensuremath{r_\text{min}}}
\newcommand{\rmax}{\ensuremath{r_\text{max}}}
\newcommand{\vel}{{\varv}}
\newcommand{\vinfty}{\ensuremath{\vel_\infty}}
\newcommand{\vmin}{\ensuremath{\vel_\text{min}}}
\newcommand{\vdis}{\ensuremath{\vel_\text{dis}}}
\newcommand{\nbcl}{\ensuremath{N_\text{cl}}}
\newcommand{\ncl}{\ensuremath{n_\text{cl}}}
\newcommand{\rhosw}{\ensuremath{\rho_\text{sw}}}
\newcommand{\rhoicm}{\ensuremath{\rho_\text{ic}}}
\newcommand{\rhocl}{\ensuremath{\rho_\text{cl}}}
\newcommand{\rhocli}{\ensuremath{{\rhocl}_i}}
\newcommand{\vcl}{\ensuremath{V_\text{cl}}} 
\newcommand{\vcli}{\ensuremath{{\vcl}_i}} 
 
\newcommand{\vsw}{\ensuremath{V_\text{w}}} 
\newcommand{\mcl}{\ensuremath{m_\text{cl}}} 
\newcommand{\mcli}{\ensuremath{{\mcl}_i}} 
\newcommand{\micm}{\ensuremath{m_\text{ic}}} 
\newcommand{\msw}{\ensuremath{m_\text{w}}} 
\newcommand{\xband}{\ensuremath{x_\text{band}}}
\newcommand{\xmin}{\ensuremath{x_\text{min}}}
\newcommand{\xmax}{\ensuremath{x_\text{max}}}
\newcommand{\dsep}{\ensuremath{d_\text{sep}}}
\newcommand{\xobs}{\ensuremath{x^\text{obs}}}
\newcommand{\xcmf}{\ensuremath{x^\text{cmf}}}
\newcommand{\nbin}{\ensuremath{N_\text{bin}}}
\newcommand{\xbin}{\ensuremath{x_\text{bin}}}
\newcommand{\doppw}{\ensuremath{\Delta\nu_\mathrm{s}}}
\newcommand{\veld}{\ensuremath{\vel_\mathrm{D}}}
\newcommand{\vref}{\ensuremath{\vel_\mathrm{s}}}
\newcommand{\ric}{\ensuremath{r^\mathrm{c}_{i}}}

\allowdisplaybreaks

\begin{document}

\title{3-D radiative transfer in clumped hot star winds}
\subtitle{I. Influence of clumping on the resonance line formation}

\author{B. \v{S}urlan\inst{1, 2, 3}
\and W.-R. Hamann\inst{4}
\and J. Kub\'at\inst{1}
\and L. M. Oskinova\inst{4}
\and A. Feldmeier\inst{4}
}
\authorrunning{B. \v{S}urlan et al.}

\institute{Astronomick\'y \'ustav, Akademie v\v{e}d \v{C}esk\'e Republiky, 
           CZ-251 65 Ond\v{r}ejov, Czech Republic
	   \and           
           Matematicko fyzik\'aln\'{\i} fakulta, Univerzita Karlova, Praha, Czech Republic
	   \and  
           Matemati\v{c}ki Institut SANU, Kneza Mihaila 36, 11001 Beograd, Republic of Serbia
	   \and
	   Institut f\"ur Physik und Astronomie, Universit\"at Potsdam, 
           Karl-Liebknecht-Stra{\ss}e 24/25, 14476 Potsdam-Golm, Germany}

\date{Received: }

\abstract{The true mass-loss rates from massive stars are important 
for many branches of astrophysics. For the correct modeling of the resonance 
lines, which are among the key diagnostics of stellar mass-loss, 
the stellar wind clumping turned out to be very important. In order to 
incorporate clumping into radiative transfer calculation, 3-D 
models are required. Various properties of the clumps may 
have strong impact on the resonance line formation and, therefore, 
on the determination of empirical mass-loss rates.} 
{We incorporate the 3-D nature of the stellar wind clumping 
into radiative transfer calculations and investigate how 
different model parameters influence the resonance line formation.} 
{We develop a full 3-D Monte Carlo radiative transfer code for 
inhomogeneous expanding stellar winds. The number density of 
clumps follows the mass conservation. For the first time, 
realistic 3-D models that describe the dense as well as the 
tenuous wind components are used to model the formation of 
resonance lines in a clumped stellar wind. At the same time, 
non-monotonic velocity fields are accounted for.}
{The 3-D density and velocity wind inhomogeneities show very strong 
impact on the resonance line formation. The  different parameters 
describing the clumping and the velocity field results in 
different line strengths and profiles. We present a set of 
representative models for various sets of model parameters and 
investigate how the resonance lines are affected. Our 3-D models 
show that the line opacity is reduced for larger clump separation 
and for more shallow velocity gradients within the clumps.}
{Our new model demonstrates that to obtain empirically correct 
mass-loss rates from the UV resonance lines, the wind 
clumping and its 3-D nature must be taken into account. 
  
\keywords{stars: winds, outflows, clumping -- stars: mass-loss -- 
stars: early-type}

}

\maketitle

\section{Introduction} 

Hot massive stars lose mass via stellar winds. Mass loss plays an 
important role in the stellar evolution and strongly affect the 
interstellar environment \citep[]{ber2008}. In order to better 
understand the evolution of hot massive stars and their influence on 
the environment it is necessary to determine their mass-loss rates 
reliably. Many sophisticated models of the stellar evolution have been 
calculated \citep[see][for a review]{maed2010}, however relying on 
mass-loss rates, that are still far from being empirically established.

Since the 1970s, significant progress in understanding the physics of 
hot star winds has been made \citep[see e.g.][]{lucy1970, cak1975, paul1986}, 
assuming the standard wind model (stationary, spherically symmetric 
wind with uniform flow). Winds of hot stars are driven by radiation 
absorbed in spectral lines, so-called the line-driven winds. For recent 
reviews of the physical mechanism of line driving and of line-driven 
stellar winds see, e.g., \cite{krt2007}, \cite{puls2008}, or 
\cite{owo2010}. 

In the last decades there was a growing evidence that the stellar 
winds are not smooth \citep[]{hamann2008}. Detailed theoretical 
studies showed that the line-driven winds are intrinsically unstable 
\citep[]{lucy1980}. Due to this instability shocks and wind density 
structures (clumps) develop in the winds. Theoretical evidence of 
clumping is based on numerical simulations of line-driven stellar winds 
that have been performed using the simplifying assumption of spherical 
symmetry and 1-D geometry \citep {feld1995, feld1997, owo1988, run2002} 
or pseudo-2D geometry \citep{dess2003, dess2005}. 

Theoretical predictions are supported by direct observational evidence 
of clumping. \cite{eve1998} found stochastic variable structures in 
the He emission line of $\zeta$ Puppis, which drift with time from the 
line centre to the blue edge of the line. These structures were explained 
by accelerated wind inhomogeneities moving outwards. High-resolution 
spectroscopic monitoring of the line-profile variations (LPVs) in 
the emission lines of nine Wolf-Rayet (WR) stars \citep{lep1999}, 
investigation of the LPVs of {\Halpha} for a large sample of O-type 
supergiants \citep{mark2005}, and direct spectroscopic observation of 
five O-type massive star of different evolutionary stages \citep{lep2008}, 
suggest that clumpy structures are a common property and a universal 
phenomenon of all hot star winds.

In addition to the stochastic small-scale wind structures, there is also
strong evidence for the presence of large-scale wind structures, namely 
the discrete absorption components (DACs). These DACs are observed to 
propagate bluewards through the UV resonance line profiles of nearly all 
O-type stars \citep{prinja1986, hamann2001}. To explain the observed DAC 
properties qualitatively, the model of Corotating Interaction Regions 
(CIR) has been proposed \citep[]{mulcir}. The CIRs form in a rotating 
stars when high-density, low-speed wind streams collide with low-density, 
high-speed streams \citep[see, e.g.,][]{cran1996}.

One of the indirect evidences of clumping in hot star winds comes from 
X-rays observations. Detailed investigation of the X-ray transport in 
clumped winds showed that wind clumping may strongly affect the X-ray 
line formation \citep[see e.g.][]{feld2003, lida2004, lida2006, owo2006}.

The radiative transfer models for non-LTE stellar-winds, like
CMFGEN \citep{hill1998}, PoWR \citep{hamann2004} and FASTWIND \citep{puls2005}, 
accaunt for wind inhomogeneities in an approximative way. An adjustable 
parameter, like the ``clumping factor'' $D$, defines the enhancement of 
the density inside clumps compared to a smooth model with the same 
mass-loss rate. Clumps are assumed to be optically thin at all frequencies, 
and the velocity field is monotonic. The empirical mass-loss rates 
derived under these assumptions (the so-called microclumping approach) 
differ from those obtained when different diagnostics are used. 
It was shown that the mass-loss rates derived from $\rho^2$-based 
diagnostics (i.e. recombination lines such as \Halpha, IR, and radio 
emission lines) have to be reduced by a factor of $\sqrt{D}$ compared 
to the values obtained under the assumption of a smooth wind 
\citep[see, e.g.,][]{hamann1998, bou2003, bou2005}. On the other hand, 
diagnostics that depend linearly on the density (e.g., unsaturated UV 
resonance lines) should not be affected by optically thin clumping. 
\cite{ful2006} found that mass-loss rates obtained from the \ion{P}{v} 
resonance lines are systematically smaller than those derived from {\Halpha}  
or radio free-free emission. Consequently, a reduction of the empirical 
mass-loss rates by up to a factor of 100 compared to the unclumped models 
was suggested. Another possibility to explain the discrepancies between 
$\rho$- and $\rho^2$-based diagnostics was proposed by \cite{wal2010}. 
They suggested that the XUV radiation near the \ion{He}{ii} ionization 
edge originating in wind shocks may destroy the \ion{P}{v} ions and, 
consequently, \ion{P}{v} resonance line diagnostic may be explained without 
the need to decrease the mass-loss rates. However, it was shown by 
\cite{nlte3} that X-rays fail to destroy the \ion{P}{v} ions.

In order to reconcile results obtained from different diagnostics,
traditionally used assumptions have to be relaxed. The first attempt in 
this direction was made by \cite{lida2007}. They studied resonance and 
recombination lines assuming statistical properties of clumps and keeping a
monotonic velocity field. They proposed that the discrepancies between 
mass-loss rates derived from recombination lines and from the \ion{P}{v}
resonance doublet can be solved by accounting for macroclumping. 
In this approach clumps can be of any optical thickness. They showed
that accounting for macroclumping has a significant impact on the line 
formation process, manifested as reduction of the effective opacity of 
the medium leading to weaker lines for a given mass-loss rate.

A further relaxation of the traditional assumptions was made by 
\cite{owo2008}. He pointed out that for line transitions, the 
non-monotonic velocity field (``vorosity'') can be important. He 
showed that a non-monotonic velocity field may also reduce the 
effective opacity. \cite{zsa2008} stressed that the void inter-clump 
medium (ICM) assumption also has to be relaxed, and that the ICM must be 
taken into account. \cite{sun2010} showed that the detailed density 
structure, the non-void ICM, and non-monotonic velocity field improve 
the line fits. They confirmed the findings of \cite{lida2007} that 
the microclumping approximation is not adequate for UV resonance line 
formation in typical OB-star winds. This is in agreement with 
\cite{prinja2010}, who established spectroscopic evidence for optically 
thick clumps in the wind by measuring the ratios of the radial optical 
depths of the red and blue components of the \ion{Si}{iv} doublets of 
B0 to B5 supergiants \citep[see also][]{massa2008}. They showed that 
these ratios are spread between 1 and 2. Since they differ from the 
predicted value of two (this value follows from the atomic constants 
for a smooth wind), this is a direct signature of optically thick 
clumping. 

In addition to the statistical approach, the Monte Carlo (MC) technique 
proved to be a very suitable method for studying clumps in hot star 
winds. \cite{lida2004, lida2006} used this approach in 2D geometry 
to calculate X-ray line profiles. In a recent paper, \cite{sun2010} 
synthesized UV resonance lines from inhomogeneous pseudo-2D 
radiation-hydrodynamic wind models and 2-D stochastic wind models 
using MC radiative transfer calculation. In their second paper, 
\cite{sun2011} extended the wind model to pseudo-3-D. 
\cite{mui2011} also used a MC method combined with non-LTE model 
atmospheres to compute the effect of clumping and porosity on the 
momentum transfer from the radiation field to the wind. They
parameterized clumping and porosity by heuristic prescriptions.

Important steps has been made in the last five years in understanding 
the line formation in clumped winds. However, all previous works 
applied simplifications to the wind geometry. The structured stellar winds
are essentially a 3-D problem, and a full description requires a 3-D 
radiative transfer. In this paper we treat for the first time the  
full problem with 3-D radiative transfer in the clumped wind. We 
accounting for: non-monotonic velocity, non-void ICM, and full 
3-D without any limitation for geometry. We present 
the basic concept of the model and the method we developed. The main 
effects on the resonance lines (both singlets and doublets) including 
wind clumping in density and velocity as well as the effect of a 
non-void ICM are demonstrated. In a forthcoming publication we will 
apply our model to observations and derive mass-loss rates.

In Sect.~\ref{wmodel} we define the wind model (geometry, velocity, and 
opacity of the wind) and our parametrization of the clump properties. 
The MC radiative transfer code is described in Sect.~\ref{montecarlo}.
Results of the model calculation are presented in Sect.~\ref{rescalc}.
In Sect.~\ref{zaver} we summarize our results and outline further work.

\section{The wind model} 
\label{wmodel}

\begin{figure}
\begin{center}
\includegraphics[width=0.65\hsize]{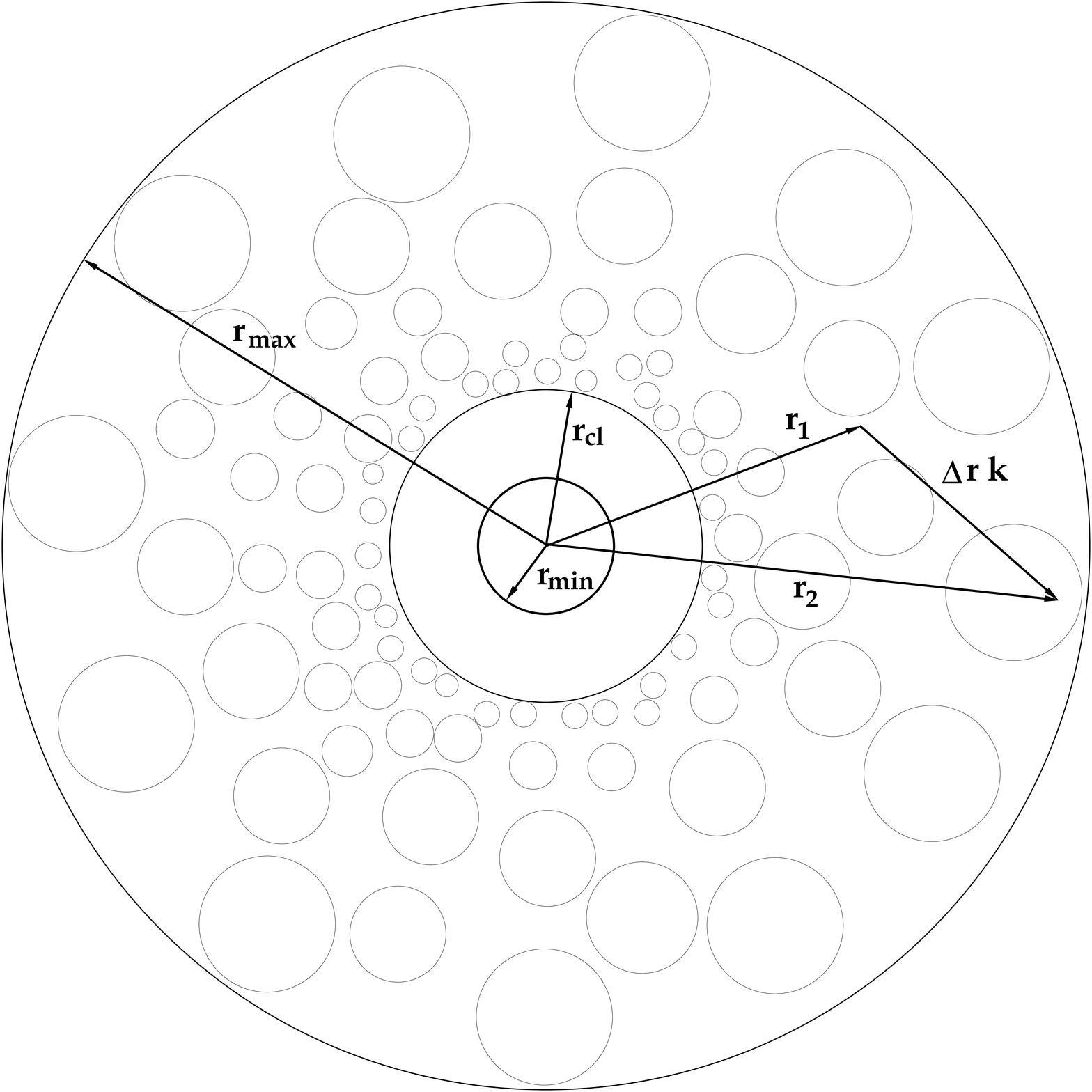}
\caption{A schematic view of the wind model. $\rmin$ is the lower 
boundary of the wind (the surface of the star), $\rcl$ is the onset 
radius of the wind clumping, and $\rmax$ is the outer boundary of 
the calculation. Vectors $\vec{r_{1}}$, $\vec{r_{2}}$, and $\vec{k}$ 
are described in Sect.~\ref{opthlou}.}
\label{fig:position-vector}
\end{center}
\end{figure}

We assumed a wind that may consist of a smooth and a clumped region. 
The clumped region comprises two density components: the tenuous ICM 
and the dense clumps. All distances in the wind are expressed in
units of stellar radius. We introduce a parameter $\rcl$ where the 
wind clumping sets on. The lower boundary of the wind $\rmin$ is set 
to the surface of the star ($\rmin=1$). The region $1\le r<\rcl$,  
represents the smooth wind region. The region $\rcl<r \le \rmax$, 
where $\rmax$ is the outer boundary of the wind, represents the 
clumped region (see Fig.~\ref{fig:position-vector}).

\paragraph{\bf General method.} 

In order to solve the radiative transfer through the clumped wind, 
we first generate a snapshot of the clumps' distribution. Using the 
MC approach (see Sect.~\ref{montecarlo}), we then follow the photons 
along their paths. The density and velocity of the wind can be
arbitrarily defined in a 3-D space. The calculations are carried out 
in the comoving frame following the prescriptions by 
\cite{hamann1980}. 

We solve the radiative transfer in the dimensionless form. We 
introduce the dimensionless frequency in the observer's frame 
$\xobs$, measured relative to the line center in units of standard 
Doppler-width \doppw,
\begin{equation}
\label{xfr}
\xobs=\zav{\frac{\nu}{\nu_{0}}-1}\frac{c}{\vel_{s}},
\end{equation}
where $\vref$ is an arbitrary reference velocity and 
$\vref=\Delta \nu_{s}(c/\nu_{0})$. During the calculations we 
strictly keep the local co-moving frame frequency $\xcmf$ of the 
photon which means that the frequency of the followed photon at 
the certain coordinate point can be expressed using the scalar 
product of the vector of the local macroscopic velocity 
($\vec{\vel}$) and the unit vector in the direction of the photon's 
propagation ($\vec{k}$),
\begin{equation}
\xcmf = \xobs - \vec{k} \cdot \vec{\vel}.
\end{equation}
Only when the photon exits the wind, $\xcmf$ must be transformed 
into the observer's frame by
\begin{equation}
\xobs=\xcmf+\vec{k} \cdot \vec{\vel}.
\end{equation}
Note that all velocities are dimensionless and measured in
units of $\vref$. 

The code which we developed is gridless and does not require
any symmetry. Instead of using a predefined grid, we introduce
an adaptive integration step $\Delta r$. After choosing $\Delta r$, 
the velocity and opacity are calculated along the photon's path.
This allows us to account for arbitrary density and velocity 
inhomogenites (for a detailed description see Sect.~\ref{opthlou}).

For the vectors we use a Cartesian coordinate system and in this 
coordinate system they are expressed by a set of coordinates 
($x$, $y$, $z$).

\paragraph{\bf Basic assumptions.} 

In order to study the basic effects of clumping on the resonance line 
formation (both singlets and doublets), we adopt a core-halo model. 
Only the line opacity is taken into account while the continuum opacity 
is neglected in the wind. We consider only pure scattering and assume 
complete redistribution. Only Doppler broadening is considered.

\paragraph{\bf Wind velocity.} 

The velocity field can be arbitrary. For simplicity, we assume that
the velocity field is radial and for the smooth wind we adopt the 
standard $\beta$-velocity law,
\begin{equation}
\vel_{r}=\vinfty \left (1-\frac{b}{r} \right)^{\beta},
\label{bvel}
\end{equation}
where $\vel_r$ is the radial component of the velocity vector,
$\vinfty$ is the wind terminal velocity and $b$ is chosen such that 
$\vmin\le10^{-3}\vinfty$.

The derivative of the Eq.~\eqref{bvel} gives radial velocity gradient
\begin{equation}
\vel^{\prime}=\deriv{\vel_{r}}{r}=\vel_{r}\,{\bf \beta}\,\frac{b}{r(r-b)},
\label{bvel2}
\end{equation}
that is important for choosing an adequate integration step.

\paragraph{\bf Wind opacity.} 

Line opacity calculation in the moving medium requires that we take 
into account the Doppler shift, which makes the opacity anisotropic 
in the observers frame. We do not adopt the Sobolev approximation but 
solve the radiative transfer equation in the comoving frame. Since 
in this frame the fluid is at rest, the opacity is isotropic.

For the opacity we use the same parameterization as \cite{hamann1980},
\begin{equation}
\chi(r)=\frac{\chi_{0}}{r^{2}\frac{\vel_{r}}{\vinfty}}\, q(r) \, \phi_{x},
\label{opa}
\end{equation}
where $\chi_{0}$ is a free parameter of the model that corresponds 
to the line strength and it is proportional to the mass-loss rate and
abundance of the absorbing ion, $q(r)$ reflects the depth-dependent
degree of the ionization of the absorbing ion and for simplicity
it is chosen to be $q(r)\equiv 1$ here, i.e. constant ionization 
condition. The absorption profile $\phi_{x}$ for the singlet lines 
is assumed to be Gaussian 
\begin{equation}
\phi_{x}=\frac{1}{\sqrt{\pi}}\,e^{-x^{2}},
\end{equation}
where $x$ is the dimensionless frequency given by Eq.~\eqref{xfr}. 
We imply that the reference velocity $\vel_{s}$ is the 
Doppler-broadening velocity $\vel_{D}$ (i.e. $\vel_{s}=\vel_{D}$) 
that includes contribution from both thermal broadening and 
microturbulence and it is constant over the whole wind.

In this parametric formalism, the smooth wind opacity $\chi(r)$
is proportional to the smooth wind density $\rhosw(r)$. In our 
calculations, we do not directly calculate wind density, but the 
depth-dependent line opacity $\chi(r)$, which is enough for 
calculation of the emergent radiation from the wind.

\subsection{Description of clumping} 

\begin{figure*}
\begin{center}
\includegraphics[width=0.4\textwidth]{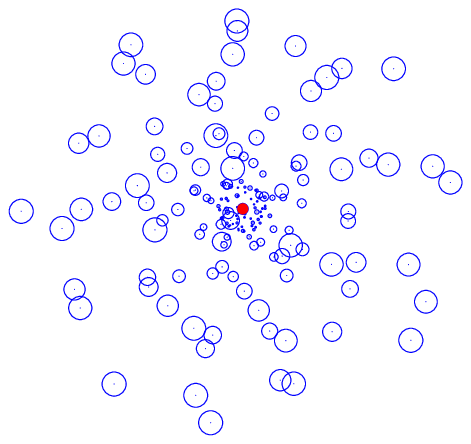} \hspace{10pt}
\includegraphics[width=0.4\textwidth]{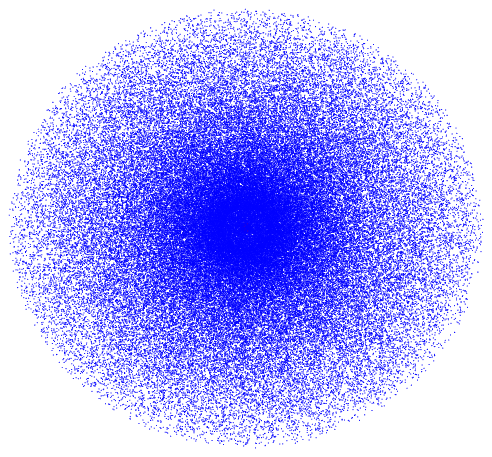} 
\caption{2-D projection of an example of a realisation of our stochastic
3-D wind model. {\bf Left:} Clumps (blue circles) with the different size 
distributed around the star (red filled circle). {\bf Right:} Distribution 
of the clumps; the filled red sphere in the center represent the star,  
while the blue dots represent the positions of the clump centers.}
\label{fig:clumps-position}
\end{center}
\end{figure*}

\subsubsection{Clump properties} %

Clumps are gaseous regions in the wind with higher density than their
surroundings, and, possibly, also with velocities different from the 
smooth wind.

We allow for arbitrary optical depth of clumps. The clumps can be 
optically thick in the cores of a resonance lines, while they may
remain optically thin at all other frequencies. Clumps are assumed 
to be stochastically distributed within the stellar wind. The average 
clump separation $L(r)$, measured between their centers, is variable 
with the distance $r$ from the star. For simplicity, we assume that 
each clump has the spherical shape with the volume $(4\pi/3)l^{3}$, 
where $l$ is the radius of the particular clump. Clump radius varies 
with the distance from the star, $l=l(r)$. The density $\rhocl$ inside 
the clump at a distance $r$ from the center is assumed to be by a 
factor $D$ higher than the smooth wind density $\rhosw$ at the same 
radius,
\begin{equation}
\label{Ddef}
\rhocl(r)=D\,\rhosw(r),
\end{equation}
and $D\ge 1$. The factor $D$ (density contrast) is the first free 
parameter of our model and for simplicity it is assumed to be depth 
independent.
Using $L$ and $l$, the factor $D$ can be expressed as
\begin{equation}
D=\frac{L^{3}(r)}{\frac{4\,\pi}{3}l^{3}(r)}.
\label{D}
\end{equation}
In our model clumps are assumed to be preserved entities. We assume 
that clumps are accelerated radially and neither split nor merge. 
The consequence of this is that the number of clumps per unit volume 
$\ncl$ has to satisfy the equation of the continuity,
\begin{equation}
\ncl\propto\frac{1}{r^{2}\vel_{r}}.
\label{nclump}
\end{equation}
Then for the average clump separation, holds $L=\ncl^{-1/3}$. With 
Eq.~\eqref{nclump}, this can be expressed as
\begin{equation}
\label{L}
L(r)=L_{0}\sqrt[3]{r^{2}\,\varw(r)},
\end{equation}
where $L_{0}$ is a second free parameter of our model that represents 
THE typical clump separation in units of the stellar radius, and 
$\varw(r)=\vel_{r}/\vinfty$ is the velocity in the units of the 
terminal speed. Similar as in the Eq.~\eqref{L}, the clump radius 
can be written as
\begin{equation}
l(r)=l_{0}\sqrt[3]{r^{2}\,\varw(r)},
\label{l0}
\end{equation}
where
\begin{equation}
l_{0}=L_{0}\sqrt[3]{\frac{3}{4\pi D}},
\label{l01}
\end{equation}
which follows from Eqs.~\eqref{D} and \eqref{L}. With this procedure 
we create clumps whose size varies with the radial distance 
(see Fig.~\ref{fig:clumps-position}).

The volume filling factor $f_{V}$ is defined as the ratio of the total 
volume that clumps occupy ($\vcl$) to the whole volume $\vsw$ in which 
the clumps are distributed (between $\rcl$ and $\rmax$),
\begin{equation}
\label{fv1}
f_{V}=\frac{\vcl}{\vsw}=\frac{\displaystyle\sum_{i=1}^{\nbcl}\vcli}
{\dfrac{4\pi}{3}(\rmax^{3}-\rcl^{3})},
\end{equation}
where $\vcli=(4\pi/3)l_i^3$ is the volume of $i$-th clump and
$\nbcl$ is the total number of the clumps.

For the case of void ICM, the entire mass of the wind
$\msw=\int_{\vsw} \rhosw(r) \,\de V = \langle\rhosw\rangle\,\vsw$
in the clumps can be written as (using Eq.~\ref{Ddef})
\begin{equation}
\label{msw}
\msw=\displaystyle\sum_{i=1}^{\nbcl}\,
\mcli=\displaystyle\sum_{i=1}^{\nbcl}\,
\rhocli\,\vcli= D\sum_{i=1}^{\nbcl} \rhosw^i\vcli
= D\,\langle\rhosw\rangle \sum_{i=1}^{\nbcl} \vcli
\end{equation}
where $\mcli$ and $\rhocli$ are mass and density of the $i$-th
clump, respectively, and $\rhosw^i$ is the value of the smooth wind
density at the location of the $i$-th clump. If the ICM is void,
\begin{equation}
\label{fvsw}
f_{V}=\frac{1}{D}.
\end{equation}

\subsubsection{Inter-clump medium} 

The ICM density $\rhoicm$ is assumed to be reduced with respect to 
the smooth wind density $\rhosw$ (the density of the wind without any 
clumping) by a factor $d$ (third free parameter of our model, assumed 
to be radius independent),
\begin{equation}
\rhoicm(r)=d\,\rhosw(r); \quad 0\le d < 1.
\end{equation}

For the case of non-void ICM, the total mass of the wind is distributed 
between clumps and ICM. In this case, the total mass of the wind
$\msw= \langle\rhosw\rangle\,\vsw$ can be expressed as
\begin{equation}
\label{msw2}
\msw=\micm + \displaystyle\sum_{i=1}^{\nbcl}\,
{\mcli} =d\,\langle\rhosw\rangle(\vsw-\vcl) +
D\,\langle\rhosw\rangle\sum_{i=1}^{\nbcl}\,{\vcli}.
\end{equation}
Then from the Eqs.~\eqref{fv1} and \eqref{msw2} follows 
\begin{equation}
f_{V}=\frac{1-d}{D-d}.
\label{fvicm}
\end{equation}

\subsubsection{Clump distribution} %

The distance $r_i$ of the $i$-th clump from the stellar center is 
chosen randomly. By employing the von Neumann rejection method 
\citep{press1992} with the inverse of the velocity law $1/\vel_{r}$
as the probability density distribution function. This reflects 
the equation of the continuity for the number density of clumps.
Consequently, more clumps are concentrated close to the star
(see Fig.~\ref{fig:clumps-position}).

Clumps are distributed uniformly in $\cos{\theta_{i}}$ and
$\varphi_{i}$, and $\theta_{i}$ and $\varphi_{i}$ are randomly chosen 
with $\cos{\theta_{i}}=\mu_{i}=\sqrt{\nahoda_{i1}}$ and 
$\varphi_{i}=2\pi\nahoda_{i2}$ ($\nahoda_{i1}$ and $\nahoda_{i2}$ are 
two different random numbers). For given $r_i$, the radius of each 
clump $l_i$ is determined according to Eq.~\eqref{l0}. We do not allow 
the clumps to overlap.

\subsubsection{Inhomogeneous velocity in clumps} %
\label{inhomvel}

\begin{figure}
\begin{center}
\includegraphics[width=1.0\hsize]{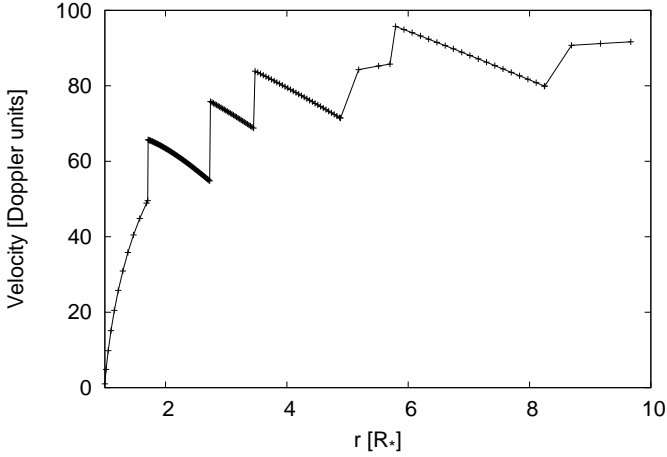}
\caption{Velocity structure of the wind model along the considered
photon path for the case of non-monotonic velocity distribution inside 
clumps.}
\label{fig:velocity}
\end{center}
\end{figure}

In our wind model the velocity of the smooth and the inter-clump 
medium is assumed to be monotonic, Eq.~\eqref{bvel}. However, the 
velocity inside the clumps is allowed to deviate from the monotonic 
wind. Consequently, a negative velocity gradient can appear 
(see more in the Sect.~\ref{montecarlo}). This assumption is based on 
the prediction of hydrodynamic wind simulations \citep[e.g.][]{owo1988, 
felpulpal, run2002}, which show that over-dense regions inside the wind 
are slower and that ICM has approximately the same velocity as the 
smooth wind. 

The velocity inside the $i$-th clump can be approximated as
\begin{equation}
\label{veldev}
\vel(r)=\vel_{\beta} (\ric) - \vdis(r) \,\frac{r- \ric}{l_{i}},
\end{equation}
where $\ric$ is the absolute position of the center of the $i$-th clump, 
$\vel_\beta(r^\mathrm{c}_{i})$ is the velocity determined according to 
Eq.~\eqref{bvel} at the position $r^\mathrm{c}_{i}$. The velocity 
dispersion $\vdis(r)=m\,\vel_{\beta}(r)$ where $m$ ($ 0 < m\le 1$) 
is the velocity deviation parameter (fourth free parameter of our model). 
Eq.~\eqref{veldev} introduces a negative velocity gradient inside 
clumps, while the center of clump has the velocity as the smooth wind. 
The velocity structure of our model along one particular photon path 
is shown in Fig.~\ref{fig:velocity}.

\section{Monte Carlo radiative transfer} 
\label{montecarlo}

The MC approach is the most promising way for treatment of radiation 
transfer in a clumped stellar wind. ``Classical'' solution methods for 
the radiative transfer equation, like the Feautrier scheme or short 
characteristics methods, become extremely time and memory consuming 
and need very sophisticated solution schemes when we go to more spatial 
dimensions than one \citep[e.g.,][]{rtm1,rtm2,lobl08}. This is not the 
case of MC methods, where the extension to 2-D or 3-D is relatively 
simple.

\subsection{Creation of a photon} 
\label{novyfoton}

A photon is released from the lower boundary of the wind (surface of 
the star) and then follow through the wind until they reach the outer 
boundary of the wind or scatters back into the photosphere. The photons 
are uniformly distributed in {\bf $\cos{\theta}$} and $\varphi$ over the 
whole surface area of the star. The photons from the lower boundary are 
released only upwards uniformly in $\varphi$ and with a distribution 
function $\propto \mu d\mu$ in $\mu$. The angular distribution 
function for photons emitted by the photosphere follows from the 
definition of the flux \citep[see, e.g.,][]{lucy1983}. The initial 
unit vector $\vec{k}$ of the photon propagation from the surface of 
the star is randomly chosen with 
$\cos{\theta_{k}}=\mu_{k}=\sqrt{\nahoda_{k1}}$ and 
$\varphi_{k}=2\pi\nahoda_{k2}$ ($\nahoda_{k1}$ and 
$\nahoda_{k2}$ are two different random numbers).

In our MC calculations, frequencies of newly created photons are 
determined from the interval defined using the ratio 
$\vinfty/\vel_{D}$ \citep[as in][]{ham1981}. Initial frequencies of 
photons have values from the interval $\langle\xmin, \xmax\rangle$ where
\begin{equation}
\xmin=-\,\frac{\vinfty}{\vel_{D}}\,-\, \xband; \quad 
\xmax=\frac{\vinfty}{\vel_{D}}\,+\, \xband. 
\end{equation}
where $\xband$ is the line width and it is usually assumed to be 
about 4.5 Doppler units \citep[similarly as in][]{mha}. 

We divide the whole frequency interval into $\nbin$ subintervals of 
equal length, and the same number of the photons are released in each 
subinterval. Each photon obtains the frequency $\xobs_{n}(\rmin)$, randomly 
chosen from the frequency interval of $n$-th ($n=1,\dots,\nbin$) bin, given 
in the observer's frame as
\begin{equation}
\xobs_n(\rmin)=(n-\nahoda)\xbin + \xmin,
\end{equation}
where $\xbin=\zav{\xmax-\xmin}/\nbin$ is the width of the frequency bin 
(the same for all bins) and $\nahoda$ is a random number. Then the frequency 
is transformed to the comoving-frame frequency $\xcmf_{n}$.

\subsection{Optical depth calculation} %
\label{opthlou}

After its creation, each photon obtains an information how far it is 
allowed to travel before it undergoes interaction with some particle.
This distance is determined by a randomly chosen optical depth 
$\tnahoda=-\ln\nahoda$ \citep[see][]{mcah,mccnd}. Then the actual 
optical depth, which photon passes on its travel, is calculated by 
summing opacity contribution along its path, and it is checked with 
$\tnahoda$.
\begin{figure}
\begin{center}
\includegraphics[width=0.93\hsize]{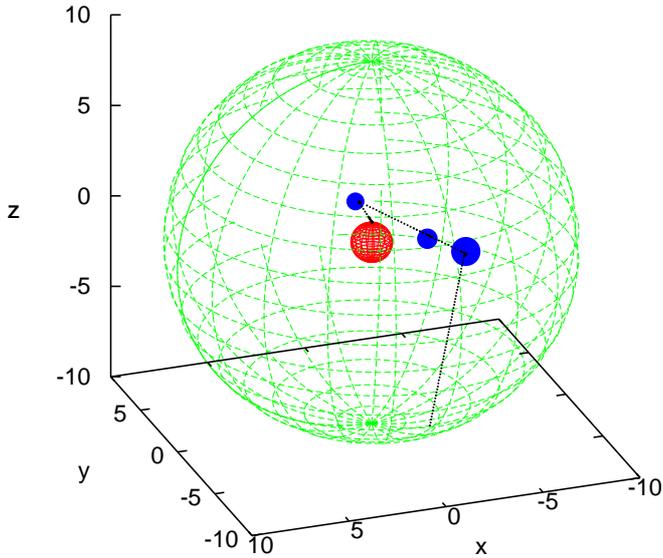}
\caption{The path of one particular photon inside a realization 
of our clumped wind with the adaptive integration step. The bigger 
green sphere represents the outer boundary while the red sphere 
in the center represents the lower boundary of the wind. Smaller 
blue balls represent the clumps and black dots denote integration 
steps.}
\label{fig:int_step}
\end{center}
\end{figure}

In order not to miss any inhomogeneities inside the wind in either 
velocity or density, we use adaptive integration steps. The photon 
path is defined by the unit vector of photon propagation $\vec{k}$ 
and the photon is followed along its path by stepwise adding a variable 
integration step $\Delta r$. If the current photon position is 
$\vec{r}_1$, then the vector of the next photon position 
$\vec{r}_{2}$ is $\vec{r}_{2}=\vec{r}_{1}\, + \, \Delta r \, \vec{k}$.
After each integration step, the coordinates of the current photon 
position are recalculated. When the photon is followed through the 
smooth part of the wind, $\Delta r$ is set not to be smaller than 
$0.1$ of the radial length of the smooth wind region and not to be 
larger than $\vel_{r}/\vel^{\prime}$, i.e., 
$0.1(\rcl-\rmin)\le\Delta r \le \vel_{r}/\vel^{\prime}$,
$\vel^\prime$ is the velocity derivative -- see Eq.~\eqref{bvel2}.
When the photon is followed inside the ICM, then 
$\Delta r = 0.3 \, l(r_{2})$, where $l(r_{2})$ is the radius of the 
clump that would have at $r_2$ (Eq.~\ref{l0}). For the photon path 
followed inside the clump, $\Delta r$ is not smaller than 
$0.1\, l(r_{2})$ and not larger than $0.3\, /\vel^{\prime}$, i.e., 
$0.1l(r_2)\le\Delta r\le 0.3\, /\vel^{\prime}$. This adaptive 
integration step technique ensures that opacity calculation along 
the photon path is done properly without a chance to skip any 
inhomogeneity inside the wind and that it is fast enough. 
The illustration of the adaptive integration step is shown in 
Fig.~\ref{fig:int_step}. 

For the optical depth calculation, it is necessary to know the 
opacity. According to Eq.~\eqref{opa}, for the opacity calculation 
at the current position we need to know $\vel_{r}$ and the Doppler 
shifted frequency of the photon at every position of the integration 
process. When the photon is traced inside the smooth or inter-clump 
regions, the velocity is calculated according to Eq.~\eqref{bvel}.
When the photon is traced inside the $i$-th clump and the option 
of the inhomogeneous velocity is turned on (Sect. \ref{inhomvel}),
then the velocity is calculated according to Eq.~\eqref{veldev}.

Along the photon path the Doppler shifted comoving-frame frequency 
of the photon is calculated. 

The optical depth between $\vec{r}_1$ and $\vec{r}_2$ for the actual 
integration step $j$ is
\begin{equation}
\tau_{j}=\int_{r_{1}}^{r_{2}}{\chi(r)dr}
\end{equation}
and it is calculated using the trapezoidal rule. Along the whole path, 
the optical depth is accumulated ($\tau=\sum_{j}^{J}{\tau_{j}}$, 
where the $J$ is the total number of the integration steps made before 
scattering happens), and when the total optical depth $\tau\ge \tnahoda$, 
then the condition for the line scattering is fulfilled.

After the scattering, the photon obtains a new direction $(\theta,\phi)$,
chosen randomly for the case of isotropic scattering as 
$\cos\theta=2\nahoda-1$ and $\phi=2\pi\nahoda$ and a new optical depth
$\tnahoda=-\ln\nahoda$ is randomly chosen again. For determination of 
$\theta$, $\phi$, and $\tnahoda$ three different random numbers are used. 
Assuming complete redistribution the photon frequency after scattering 
is calculated randomly from Gaussian distribution with the mean zero and 
the standard deviation unity \citep{box1958}. 

In order to reproduce the emergent flux from the wind, all photons that 
reach the edge of the wind are collected and put into a proper frequency 
bin. Each of the profiles shown in this paper has been calculated for one 
random configuration of clumps. This would correspond to an observation with a
short exposure time, compared to the dynamical time scale. But on the other
hand, we count the emergent photons irrespective of their direction, while
the distant observer sees the 3-D clump configuration from one specific
direction. By this averaging over all directions, we effectively obtain a
mean emergent profile, similar to what we would get from averaging over many
different random clump configurations but seen from one specific direction.

\begin{figure*}
\begin{center}
\includegraphics[width=0.49\textwidth]{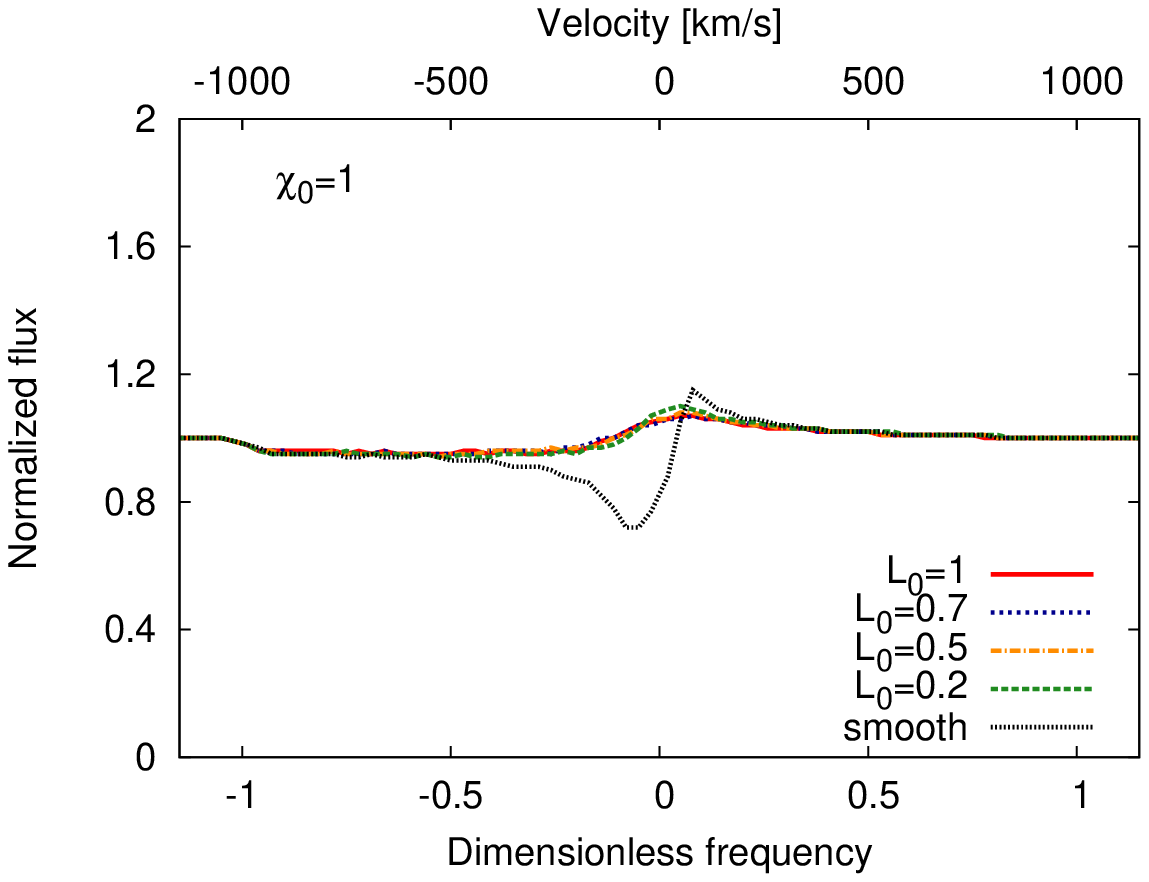}
\includegraphics[width=0.49\textwidth]{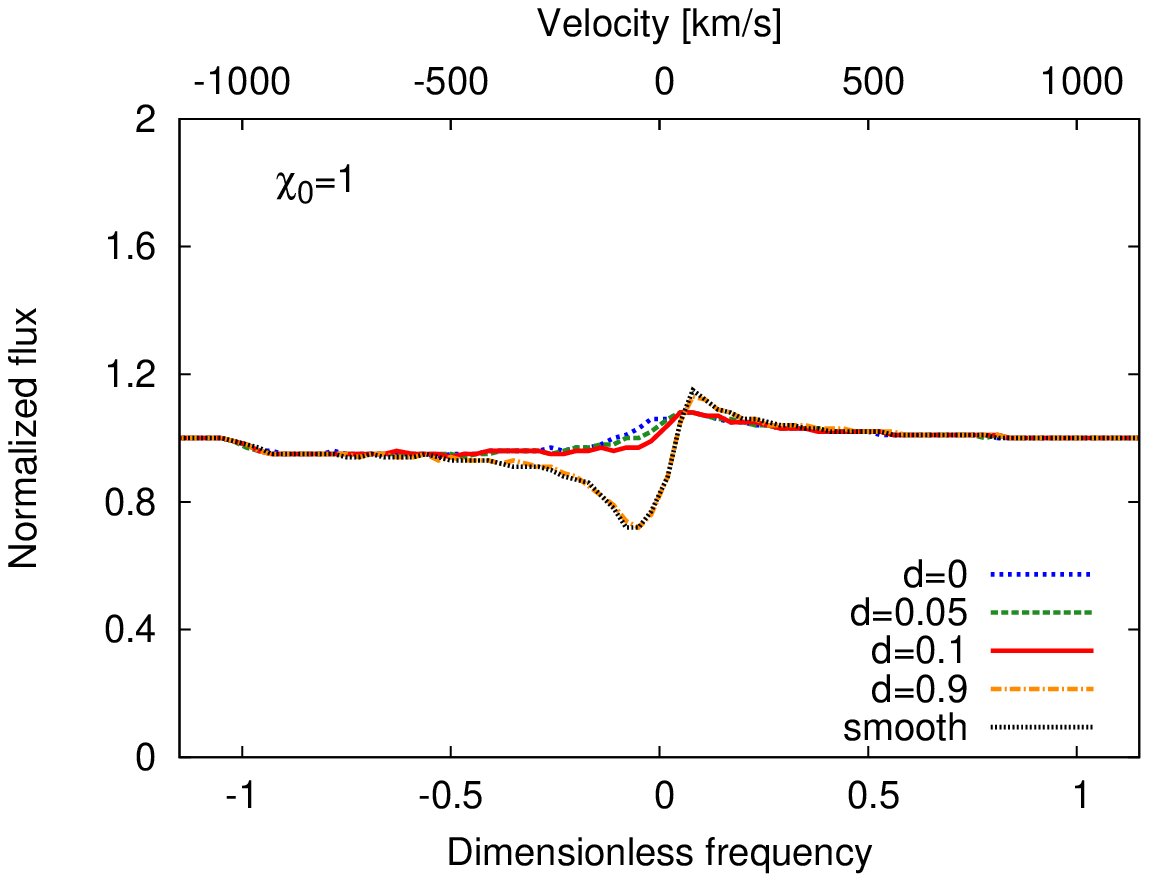}
\includegraphics[width=0.49\textwidth]{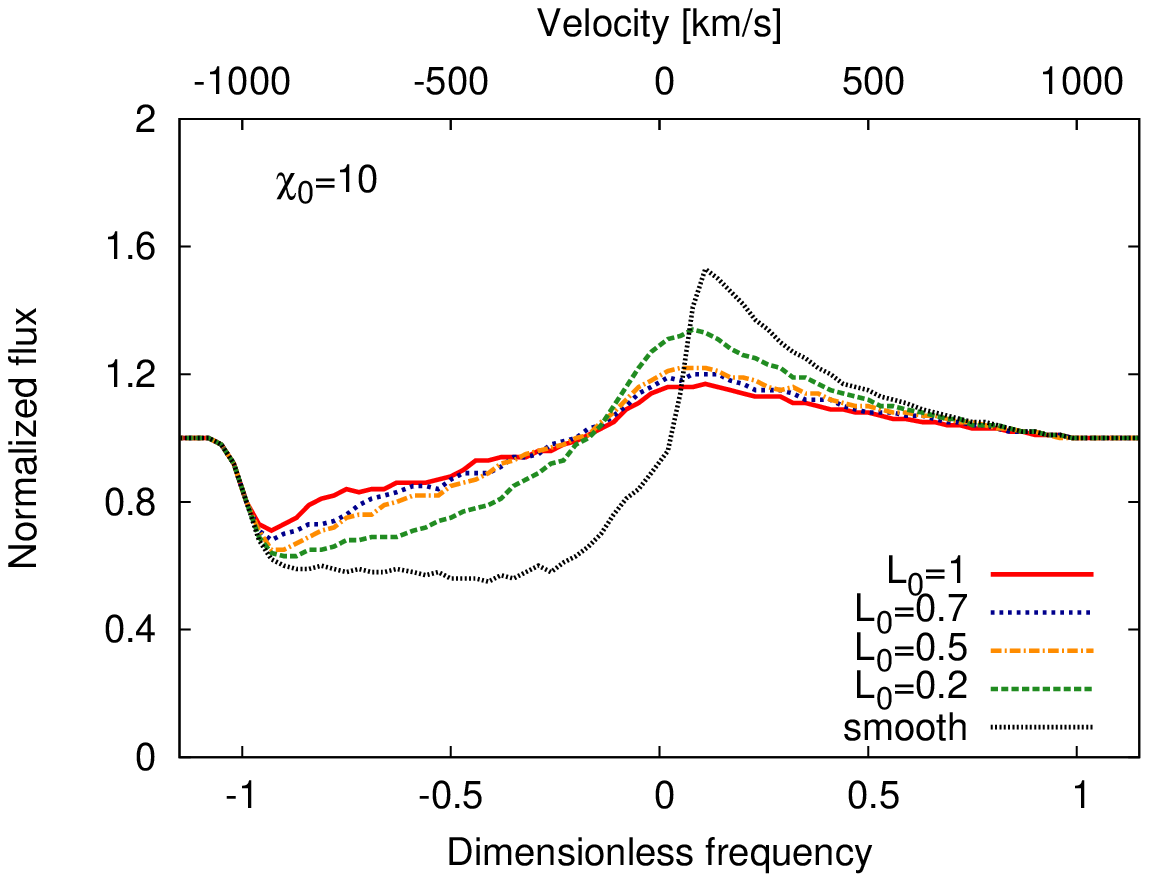}
\includegraphics[width=0.49\textwidth]{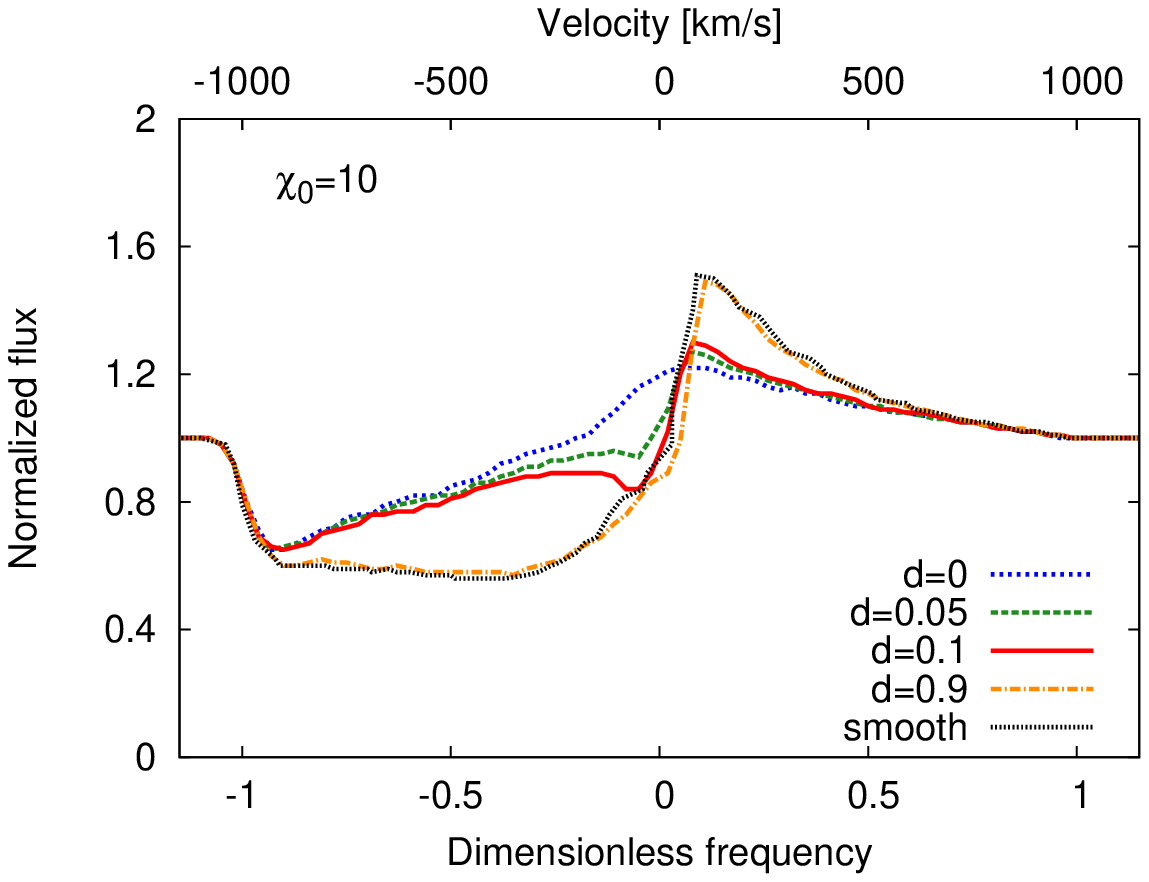}
\includegraphics[width=0.49\textwidth]{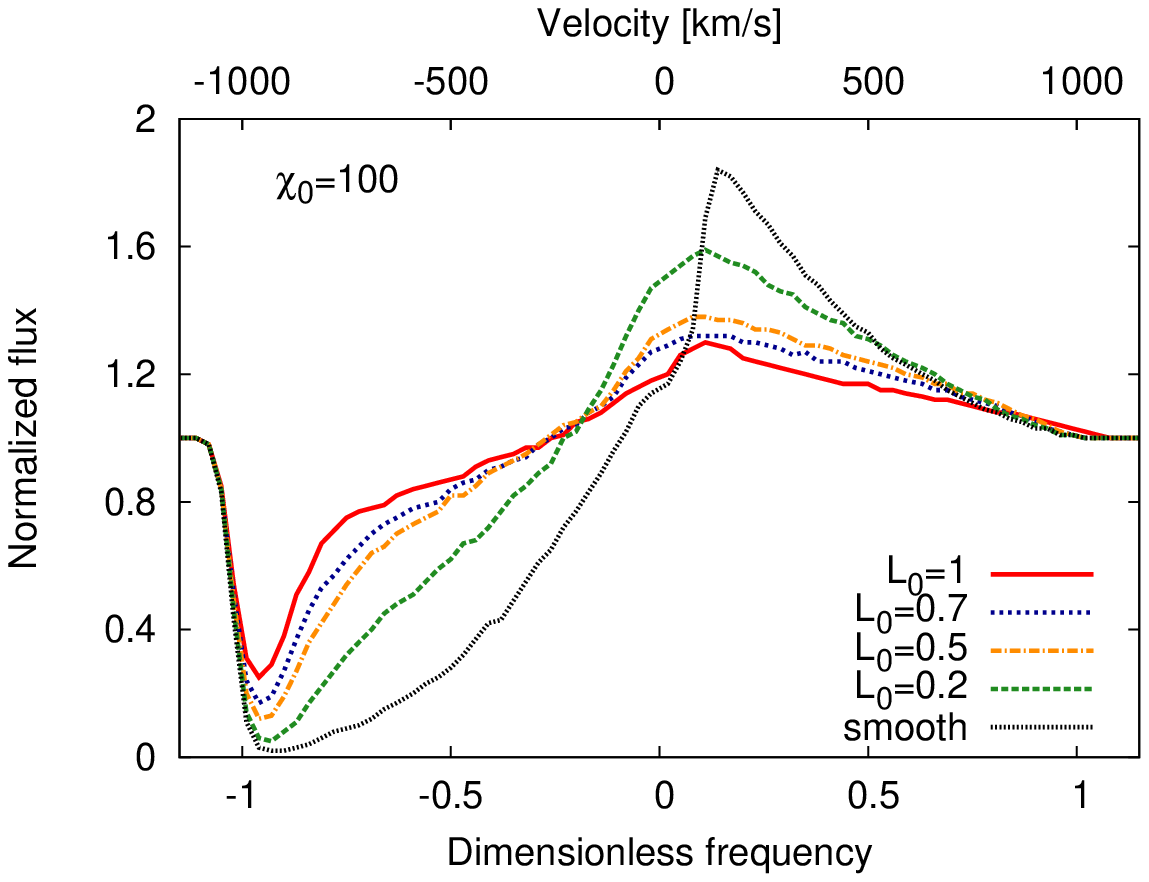}
\includegraphics[width=0.49\textwidth]{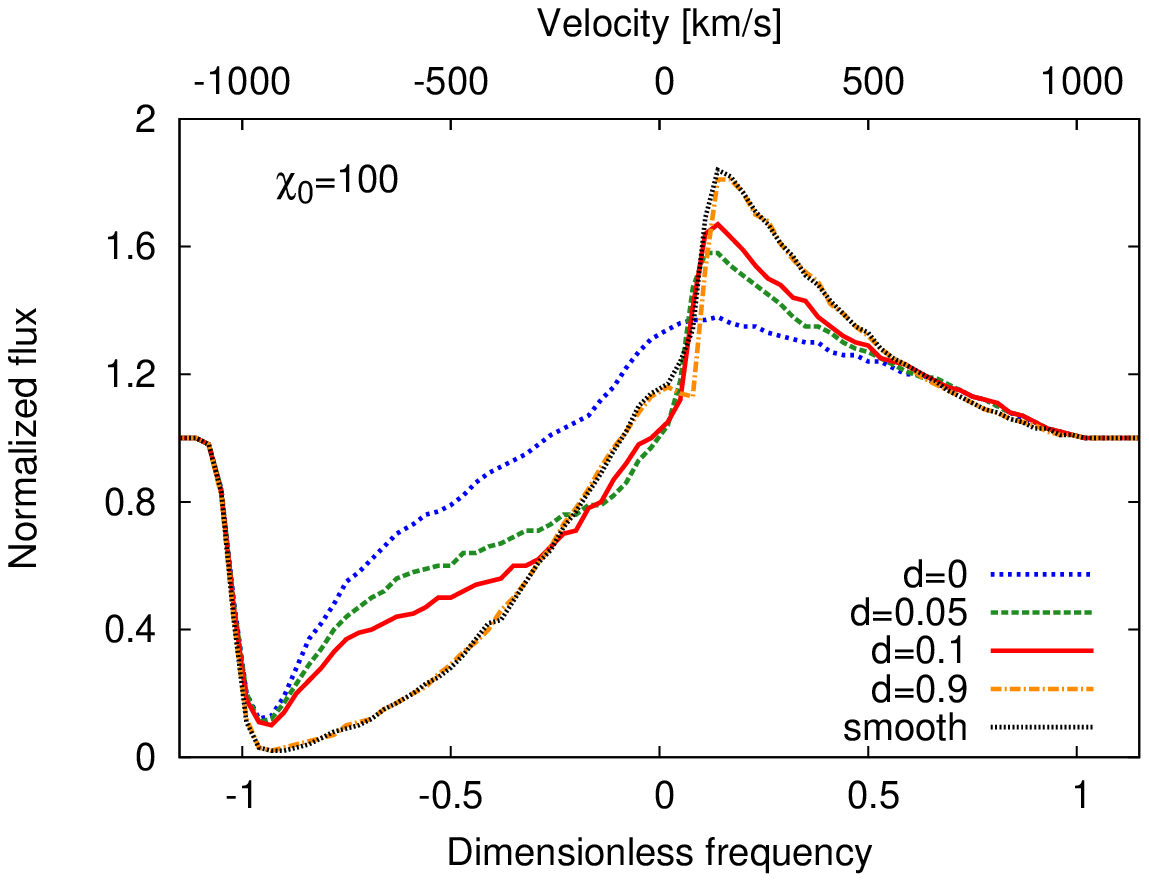}
\caption{The effects of the macroclumping (left) and the non-void 
ICM (right) on the weak ($\chi_{0}=1$, the {\em upper panels}), 
intermediate ($\chi_{0}=10$, 
the {\em middle panels}), and strong ($\chi_{0}=100$, the {\em lower 
panels}) lines. {\bf Left:} The black dashed lines represent a smooth wind 
($L_{0}\rightarrow0$) and other lines are calculated for a different 
clump separation parameter $L_{0}$ as given in the panels.
Other model parameters have their default value (Table~\ref{para-model}). 
{\bf Right:} The black dashed lines represent a smooth wind and other 
lines are calculated for different values of the ICM density 
parameter $d$ as given in the panels. Other model parameters have their 
default value (Table~\ref{para-model}).}
\label{fig:profile-L0-d}
\end{center}
\end{figure*}

\section{Results of the 3-D wind model calculation} 
\label{rescalc}

\begin{table*}
\begin{center}
\begin{tabular}{l l|l l l}
\hline  
\hline 
Fixed model  & \multicolumn {1}{c|}{Value}  &  Varied model  &   \multicolumn {1}{c}{Considered}  &  \multicolumn {1}{c}{Default}  \\
parameters   & \multicolumn {1}{c|}{}       &  parameters    &   \multicolumn {1}{c}{range}       &  \multicolumn {1}{c}{value}    \\
\hline
\hline
Outer boundary of the wind  $\rmax$ [$R_{\ast}$] &   80  &  Opacity parameter $\chi_{0}$                  &  1, 10, 100                    &   100  \\
Beta parameter $\beta$                           &    1  &  Clump separation parameter $L_{0}$            &  0.2, 0.5, 0.7, 1              &   0.5  \\
Velocity at the photosphere $\vmin$ [\kms]       &   10  &  Clumping factor $D$                           &  3, 5, 10                      &   10   \\
Terminal velocity $\vinfty$ [\kms]               & 1000  &  ICM density factor $d$                        &  0, 0.05, 0.1, 0.9             &   0    \\
                                                 &       &  Onset of clumping $\rcl$                      &  1, 1.01, 1.3, 2               &   1    \\
                                                 &       &  Doppler velocity $\vel_{D}$ [\kms]            &  20, 50, 100                   &   50   \\
       
                                                 &       &  Velocity deviation parameter $m=\vdis/\vel_{\beta}$ &  0, 0.1, 0.2             &   0    \\
\hline
\end{tabular}
\caption{The model parameters to study the influence of their variation on the resonance line profiles.}
\label{para-model}
\end{center}
\end{table*}
In this section we show basic effects on the line profiles for both 
single resonance lines and resonance doublets varying different 
model parameters. Depending on which effect we want to study, some 
parameters are kept fixed, while others are varied within considered 
ranges (see Table~\ref{para-model}). The basic model parameters 
(left part of Table~\ref{para-model}) are chosen to 
represent a typical O type star. The varied model parameters 
(the right part of Table~\ref{para-model}) have their default value 
except the selected one whose effect we aim to study. The effects 
on the line profiles are shown for weak ($\chi_{0}=1$), 
intermediate ($\chi_{0}=10$), and strong ($\chi_{0}=100$) lines.
If the variation of some model parameter shows a similar effect 
on all three types of lines, we show that effect only for the 
strong line case. 

All calculations are performed using $10^{5}$ photons distributed 
over $100$ frequency bins, resulting in a S/N ratio of about 30 per bin 
due to the Poisson statistics.

\subsection{The effects of the macroclumping} 

The total number of clumps in one snapshot of clump distribution 
can be derived using the parameter $L_{0}$ from 
\begin{equation}
\nbcl=\int\limits_{\rcl}^{\rmax}{\ncl(r)\,4 \pi\,r^{2}\, dr}.
\label{ncl}
\end{equation} 
For instance, if $0.2\le L_{0}\le 0.5$, then the total number of
clumps within the clumped region is $10^{5}\gtrsim\nbcl\gtrsim 10^{3}$.
A higher number of clumps (i.e smaller values of $L_{0}$) produces
a less porous wind. If $L_{0}\rightarrow0$ the whole wind is
filled with many little clumps and it resembles a smooth wind. 

In this subsection, all model parameters are set to their default 
values except of $\chi_{0}$ and $L_{0}$, which are varied as given 
in Table~\ref{para-model}. The main macroclumping effect on the 
line profile is the reduction of the line strength compared to the 
smooth wind. For this case, where we assumed that 
clumping starts from the surface of the star ($\rcl=0$), the variation 
of the $L_{0}$ parameter has a strong influence on all three types of 
lines (see the left part of Fig.~\ref{fig:profile-L0-d}). This is 
because the clumps are optically thick for these lines. However, 
if we assume that clumping starts at $\rcl=1.3$, the clumps 
are optically thin for the weak lines there while they remind optically 
thick for the intermediate and strong line case \citep[see][]{surlan2012}. 

According to Eqs.~\eqref{l0} and \eqref{l01}, the radius of the clumps 
depends on $L_{0}$. For very small $L_{0}$ a huge 
number of clumps with small radii are created, and the clump
contribution to the line opacity is almost the same as in case of a 
smooth wind. In this case there are not too many ``holes'' 
between the clumps, and the photons cannot escape from the wind 
easily. But for higher values of $L_{0}$, less clumps with larger 
radii exist. Consequently, there are more ``holes'' in the wind, 
through which photons may freely propagate. This leads to lower 
absorption and weaker line. 

Clumping lowers the effective opacity. This effect is weaker for the
outer parts of the wind, because the individual clumps become optically
thin there. This causes the bump in the blue part of the line. On the
other hand, a significant part of the wind has approximately the same
wind velocity close to the wind terminal velocity. Consequently, despite
lower effective opacity there is still enough matter to absorb.
This causes the absorption dip near $\vinfty$ especially for
lower $L_{0}$.

\begin{figure}
\begin{center}
\includegraphics[width=1.0\hsize]{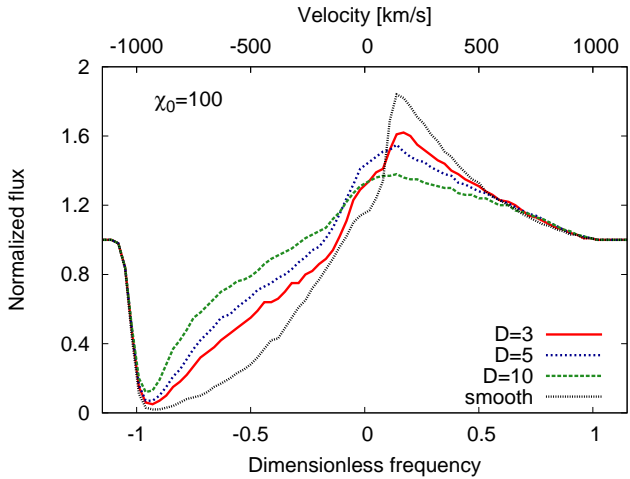}
\caption{The effect of variation of the clumping factor $D$
for the case of a strong line ($\chi_0=100$). 
The black dashed line represents a smooth wind, while the other lines 
are calculated for different $D$ as given in the figure. 
Other model parameters have their default value 
(Table~\ref{para-model}).}
\label{fig:D-k100}
\end{center}
\end{figure}

A similar effect of attenuation of the line strength can be obtained 
by varying the clumping factor $D$. In Fig.~\ref{fig:D-k100} we show
the effect only for the strong line ($\chi_{0}=100$). The model 
parameters are again set to their default values except of $D$ that is 
varied. When enhancing the density inside clumps compared to the smooth 
wind density (i.e. increasing the parameter $D$), the porosity effect 
is more pronounced and the line becomes weaker.

\begin{figure}
\begin{center}
\includegraphics[width=\hsize]{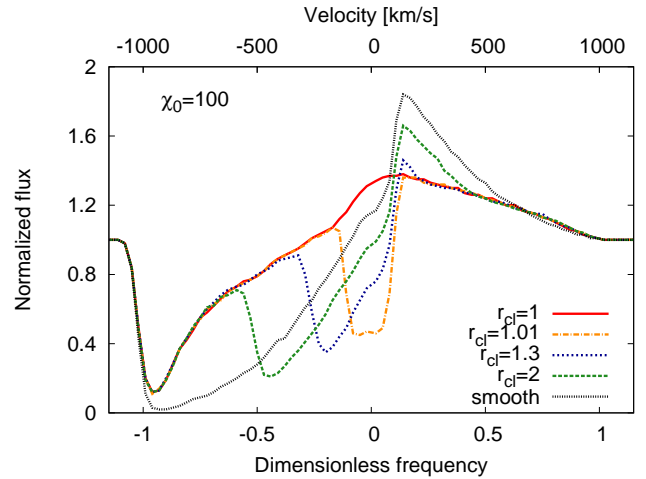}
\caption{The effect of variation of the onset of the clumping $\rcl$
for the case of strong line ($\chi_0=100$).
The dashed black line represents the smooth wind ($\rcl=\rmax$)
and other lines are calculated for different $\rcl$ as given
in the figure. Other model parameters have their default value 
(Table~\ref{para-model}).}
\label{fig:rcl}
\end{center}
\end{figure}

\subsection{Effects of the non-void inter clump medium} 

In this subsection we study the changes caused by variations of
the ICM density parameter $d$ and $\chi_{0}$, while the other model  
parameters are set to their default values. The space between clumps is 
filled with some amount of matter with density defined by $d$. This 
matter fills the density ``holes'' between the clumps, and photons can 
also be scattered there. This manifests as strengthening of both
the absorption and emission parts of the line profiles with respect 
to the model with void ICM (see the right part of 
Fig.~\ref{fig:profile-L0-d}). This effect is most pronounced for the 
strong lines, where even for a small $d$ the ICM contributes 
a lot to the absorption coefficient and makes the lines more saturated 
and the emission parts stronger (see the right lower panel in 
Fig.~\ref{fig:profile-L0-d}). For the same value of 
the parameter $d$, the strong lines are saturated more than the 
intermediate lines. A proper choice of $d$ can saturate the strong 
lines but still keep the intermediate lines unsaturated
(e.g. see lines for $d=0.1$ in the right middle and lower panels of 
Fig.~\ref{fig:profile-L0-d}).

Assuming that the ICM is not void has an influence on the strength of 
the line as a whole, but it has a particular effect on the center 
of the line, where a small absorption dip appears. Because more 
clumps are concentrated close to the star, the inner part of the wind 
together with the ICM contribute to the line opacity more than the 
outer part of the wind.

\subsection{Effects of the onset of clumping} %

In order to show the effect of the radius where clumping starts, 
we now fix all parameters of the model at their default values except
the onset radius of clumping $\rcl$. The main effect of the $\rcl$ 
variation is the appearance of a strong absorption near the line center 
(see Fig.~\ref{fig:rcl}), which is due to the smooth part of the wind 
inside $\rcl$ at low expansion velocity. If clumping starts higher in 
the wind, the absorption near the line center is broader and also 
partly attenuates the emission from the back hemisphere of the wind. 
Therefore, this central absorption dip can be used 
as diagnostic for the onset of clumping.

It is interesting to note that even a very little smooth part of the 
wind (between $r=1$ and $r=1.01$) causes significant central absorption, 
an effect which should be observable. As it can be seen from the
Fig.~\ref{fig:rcl}, only when $\rcl=1$ the absorption near the line 
center disappears. When the clumped part of the wind is larger 
(smaller value of the  $\rcl$), the reduction of the line strength 
is more pronounced. By setting $\rcl=\rmax$, the smooth wind is 
reproduced.

\subsection{Effects of velocity dispersion inside clumps} %

The absorption of an individual clump can be broadened 
by stochastic (thermal or micro-turbulent) motions inside the clump, 
but also by an additional velocity 
gradient in the clump as predicted by hydrodynamic simulations.
These kinds of line broadening are 
described in our model by the Doppler-broadening velocity (\veld), 
and by the velocity dispersion inside the clumps ($\vdis$), 
respectively (ses Sect.~\ref{inhomvel}). 

\begin{figure}
\includegraphics[width=\hsize]{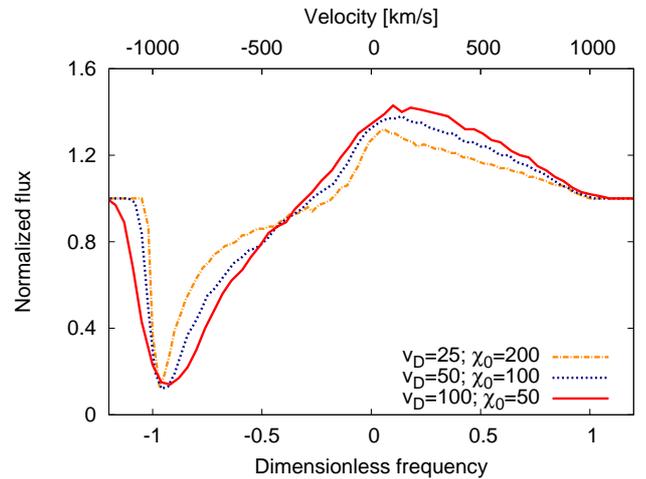}
\caption{The effect of different Doppler broadening on the line 
profiles represented by three values of $\veld$ and $\chi_0$
as given in the figure. Other model parameters have their default 
value (Table~\ref{para-model}).}
\label{fig:V_dopp}
\end{figure}

In Fig.~\ref{fig:V_dopp} we show the effect of the $\veld$ variation. 
The normalization of the frequency-integrated absorption coefficient 
is maintained by a compensating change of $\chi_{0}$. For the higher 
value of $\veld$ the line profile is broader, and absorption and 
emission are stronger. When $\veld$ decreases, the macroclumping effect 
becomes more pronounced because the clumps are optically thicker in 
the line center, which leads to a smaller effective 
optical depth due to a larger macroclumping effect. 

\begin{figure}[!htb]
\begin{center}
\includegraphics[width=\hsize]{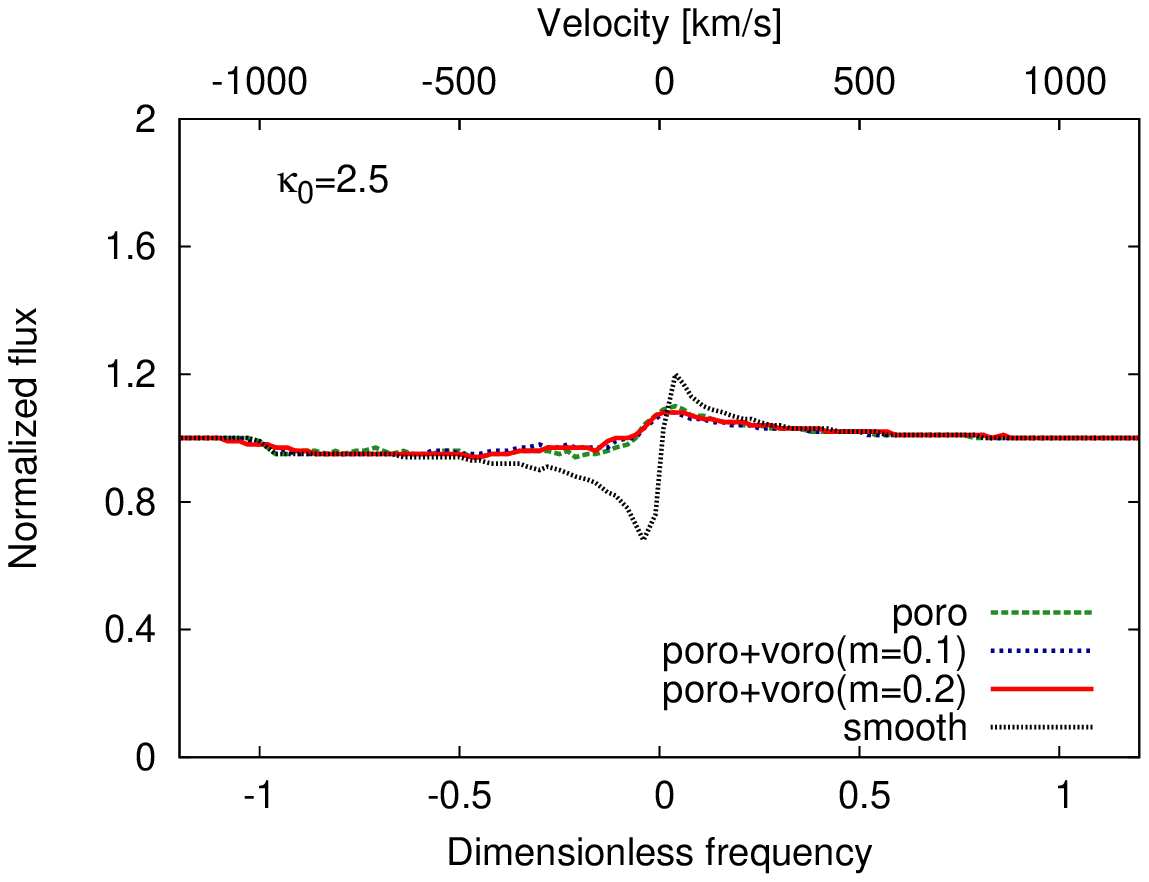}
\includegraphics[width=\hsize]{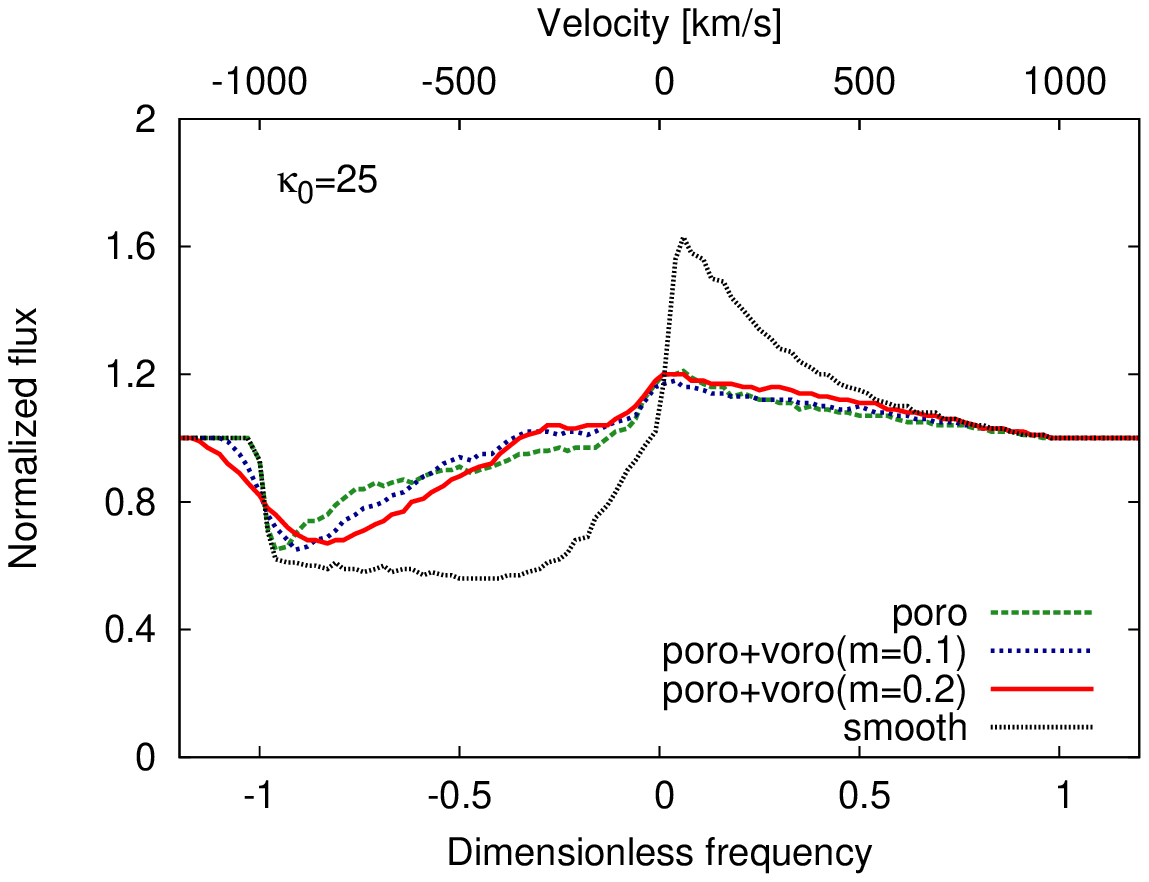}
\includegraphics[width=\hsize]{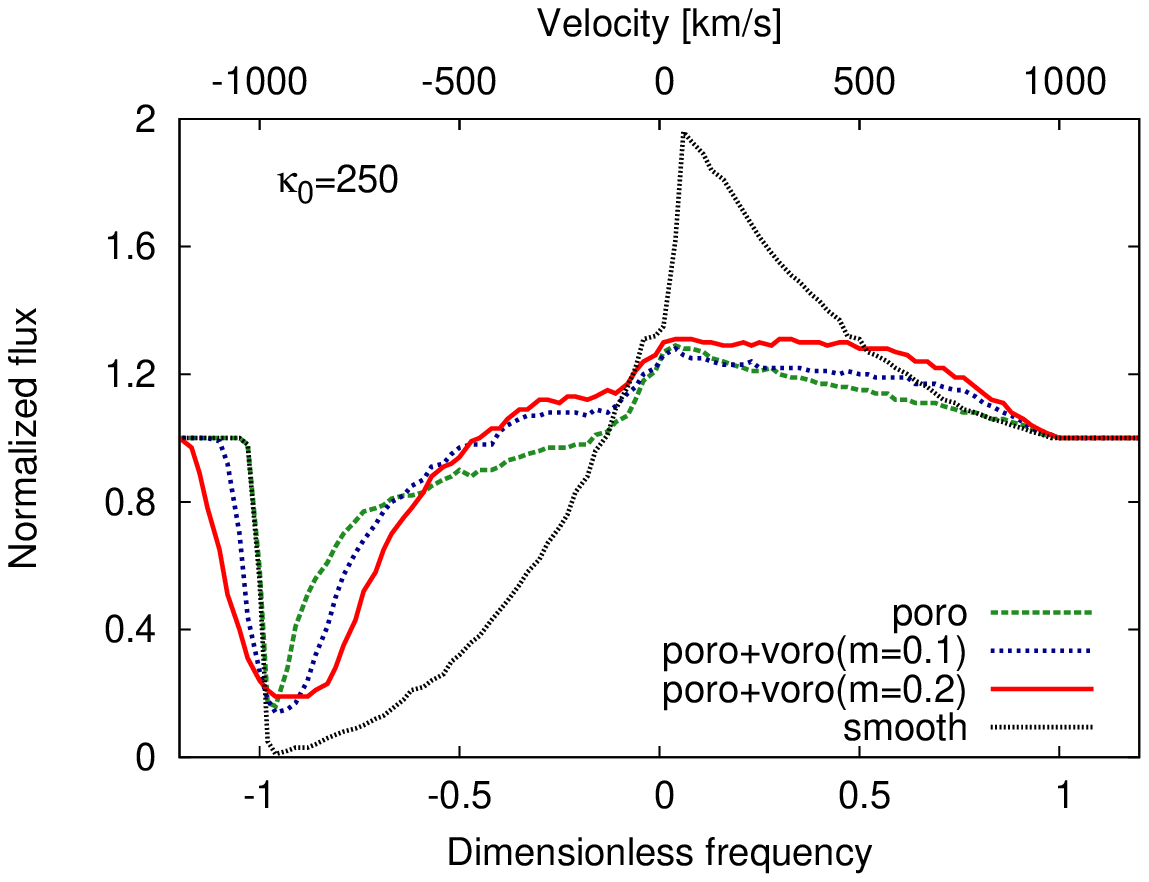}
\vspace{10pt}
\caption{Effects of the velocity dispersion inside clumps on 
the line profile. {\em Upper panel:} weak lines ($\chi_{0}=2.5$), 
{\em middle panel:} intermediate lines ($\chi_{0}=25$), 
{\em lower panel:} strong lines ($\chi_{0}=250$). The black dashed 
lines represent the smooth wind, the green dashed lines (poro) 
represent pure porous wind, the other lines (poro+voro) represent 
the porous wind with non-monotonic velocity described with $m$ as 
given in the panels. $\veld=20\,\kms$ and other model parameters have 
their default value (Table~\ref{para-model}).}
\label{fig:profile-vdev}
\end{center}
\end{figure}

To study the effect of the velocity dispersion inside clumps, we fix 
$\veld=20\,\kms$ and vary $m$ and $\chi_{0}$ (Fig.~\ref{fig:profile-vdev}). 
If the velocity dispersion inside the clumps is higher (i.e.\ when 
increasing the parameter $m$), the gaps in the velocity field are smaller 
(more velocities overlap) and the probability of photon escape is lower.
This leads to some absorption at velocities higher than $\vinfty$. However, 
in the shown example the velocity gradient inside the clumps does not differ 
much from the velocity gradient of the smooth wind except of its sign, thus 
leaving the optical depth of the individual clumps is roughly the same. 
Therefore, the ``vorosity'' has only little effect on the total line strength. 

When the velocity dispersion is accounted for, 
the absorption extends to velocities higher than the terminal velocity 
$\vinfty$. This effect is important for the derivation of $\vinfty$ from 
observations \citep[i.e. see][]{prinja1990}, and may provide an effective 
diagnostic for the existence of the velocity dispersion inside 
the clumps. 

\subsection{Doublet calculation} 
\label{doub}

For the calculation of doublets, the profile $\phi_{x}$ in 
Eq.~\eqref{opa} is assumed to have the form
\begin{equation}
\phi_{x}=\frac{1}{\sqrt{\pi}}\,( e^{-(x+\dsep/2)^{2}} + p\,e^
{-(x-\dsep/2)^{2}} ),
\end{equation}
where $\dsep$ is the separation between doublet components. The ratio of 
line opacities has a fixed value $p=\chi_{1}/\chi_{2}$, which follows 
from the atomic line strengths. The zero point frequency is located in 
the middle between the components.

The frequency of the photon after scattering is chosen randomly 
with a Gaussian distribution in the same way as for singlets. 
However, if the co-moving frame frequency 
of the photon before scattering indicates that it belongs to the 
blue component ( $\xcmf > 0$), we assume
that the photon is redistributed only within the blue component of the doublet, 
and vice versa.

\begin{figure}[!htb]
\begin{center}
\includegraphics[width=0.5\textwidth]{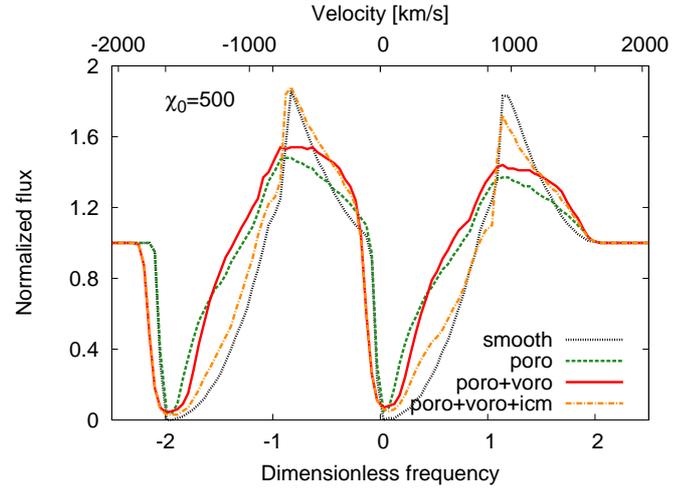}
\caption{Effects of the macroclumping on the strong doublet line 
profiles ($\chi_0=500$) including non-void ICM and velocity dispersion 
inside clumps. 
The black dashed line represents the smooth wind, the green dashed 
line (poro) represents the macroclumping effect, the full red line 
(poro+voro) represents the macroclumping effect including non-monotonic 
velocity described inside clumps with $m=0.2$, and the orange dash dotted 
line (poro+voro+icm) represents the macroclumping effect including 
non-monotonic velocity with $m=0.2$, and non-void ICM with $d=0.05$. 
Other model parameters have their default value (Table~\ref{para-model}). 
}
\label{fig:porovoro-doub}
\end{center}
\end{figure}

In Fig.~\ref{fig:porovoro-doub} we show one example of a strong doublet 
with $\dsep=2000\,\kms$ and $\chi_{0}=500$. We demonstrate  
different effects on the line profile by changing the properties of the 
clumps. These effects are analogues to the single-line case. 
The main effect of macroclumping is the reduction of the line 
strength. If the velocity dispersion inside clumps is taken into 
account, there is some absorption at velocities higher 
than $\vinfty$. Only for the case of non-void ICM it is possible 
to saturate the line. 


\section{Summary} 
\label{zaver}

In this paper we present a full 3-D inhomogenous stellar wind 
model and solve the radiative transfer to model the resonance lines. 
To our knowledge this is the first work where the problem of 
resonance line formation in stellar winds is solved in full 3-D, 
while previous work was restricted to 2-D or pseudo-3-D geometries 
only. 

The radiative transfer is also calculated for the formation of 
doublets. This is important, because UV resonance doublets are a key 
diagnostic for stellar winds. Our work demonstrates the role 
which stellar wind clumping plays in the formation of spectral lines, 
and how it affects the empirical wind diagnostics. 

The method we develop here is very flexible, as it is capable to 
account for different 3-D shapes of clumps, for arbitrary 3-D 
velocity fields, and for ICM the same time. Below we briefly 
summarize the main results of our models, deferring the 
application to observed spectra to forthcoming work 
(\v{S}urlan et al., in prep.).

\begin{itemize}
\item  Allowing for clumps of any optical depths (macroclumping) 
causes a reduction of the effective opacity in the lines.
Consequently, there is less absorption in the wind. 
Since this reduction is weaker for the outer parts of the wind,
the line profiles show an absorption dip near 
$\vel_\infty$.

\item  For a given clumping factor $D$, the key model parameter affecting the 
effective opacity is $L_0$, the clump separation parameter. The opacity 
reduction is largest for the largest $L_0$ (Fig.~\ref{fig:profile-L0-d}). 
Therefore, the mass-loss rate empirically obtained by fitting the UV 
resonance lines becomes larger when the macroclumping effect is taken 
into account.
 
\item The onset of clumping, $\rcl$, affects the line shape: the closer 
to the stellar surface clumping starts, the more pronounced is the 
absorption dip at the line center (Fig.~\ref{fig:rcl}). This absorption 
\\ \\ \\ 
dip may provide an effective a diagnostic for the onset of clumping. 

\item The line saturation is strongly affected by the ICM. 
A non-void ICM  is required to reproduce the saturated lines 
simultaneously with non-saturated lines 
(Fig.~\ref{fig:profile-L0-d}).  

\item When accounting for a velocity dispersion within the clumps, 
added to the mean velocity law, the absorption extends to a larger 
blue-shift than corresponding to $\vinfty$. This effect has to be 
taken into account when deriving $\vinfty$ from observations. 

\item In any clumped wind, non-monotonic velocities will always 
appear together with the density inhomogeneities. Therefore, their 
combined effect must be taken into account for the line formation 
modeling.

\item In case of resonance doublets, the clumping effects are 
analogues to the case of single lines. 

\end{itemize}

The main conclusion of our work is that in a realistic 3-D wind 
with density inhomogeneities and non-stationary velocity, the 
P-Cygni profiles from resonance lines are different from those from 
smooth and stationary 3-D winds. Any mass-loss diagnostics which do 
not account for wind clumping must underestimate the actual mass-loss 
rates. This can explain the reported discrepancies between the mass-loss 
rates obtained from $\rho$- and $\rho^2$-based diagnostics, respectively. 
Using our general description of clumping presented in this work, it 
will be possible to determine improved values of the stellar mass-loss 
rates.

\begin{acknowledgements} 

This work was supported by grants GA \v{C}R 205/08/0003 and 
205/08/H005, GA UK 424411, DAAD/AV\v{C}R D3-CZ2/2011-2012, and FKZ
50\,OR\,1101 (LMO). B\v{S} thanks to Ministry of Education and Science 
of Republic of Serbia who supported this work through the project 
176002 "Influence of collisions on astrophysical plasma spectra". 
WRH and LMO are very grateful for the hospitality at the Ond\v{r}ejov 
observatory. We thank the anonymous referee for useful suggestions 
regarding the clump distribution and other helpful remarks.

\end{acknowledgements}


\end{document}